\documentclass[aps,prl]{revtex4}
\usepackage{enumerate}
\usepackage{amsmath,amssymb}
\usepackage[dvips]{graphicx}
\usepackage{url} 
\usepackage{color}

\renewcommand{\textcolor}[1]{}

\def\msol{M_{\odot}}
\def\mbh{M_{\rm BH}}
\def\meff{M_{\rm eff}}

\def\rhosp{\rho_{\rm sp}}

\def\rsp{r_{\rm sp}}

\def\rmin{r_{\rm min}}

\begin{document}
\begin{flushright}RESCEU-1/13\end{flushright}
\title{
%Gravitational wave from a stellar mass and an intermediate mass black
%hole binary surrounded by a dark matter mini-spike \\
%A new signature of a dark matter mini-spike : 
%Gravitational wave from a stellar mass and an intermediate mass black
%hole binary surrounded by a dark matter mini-spike \\
%Gravitational wave as a probe of dark matter mini-spike surrounding an
%intermediate mass black hole\\
A new probe of dark matter properties: 
gravitational waves from an intermediate mass black
hole embedded in a dark matter mini-spike \\
%Gravitational wave from a dressed intermediate mass black hole \\
%New probe to a dressed black hole\\
%New probe to a dark matter mini-spike\\
%New probe to an intermediate mass black hole surrounded by a dark matter mini-spike\\
}
\author{Kazunari Eda}
\email{eda@resceu.s.u-tokyo.ac.jp}
\author{Yousuke Itoh}
\author{Sachiko Kuroyanagi}
\affiliation{
Research center for the early universe, 
School of Science, 
University of Tokyo, Tokyo, 113-0033, Japan
}
\author{Joseph Silk}
\affiliation{
Institut d' Astrophysique, UMR 7095, CNRS, UPMC Univ. Paris VI, 98 bis Boulevard Arago, Paris 75014, France
}
\affiliation{
Department of Physics and Astronomy, The Johns Hopkins University Homewood Campus, Baltimore, MD 21218, USA}
\affiliation{
Beecroft Institute for Particle Astrophysics and Cosmology, Department of Physics, University of Oxford, Keble Road, Oxford, OX1 3RH, United Kingdom
}

\begin{abstract} 
  An intermediate-mass black hole (IMBH) may have a dark-matter (DM)
  minihalo around it and develop a spiky structure within less than a parsec
  from the IMBH.  When a stellar mass object is captured by the
  minihalo, it eventually infalls into such an IMBH due to gravitational
  wave back reaction which in turn could be observed directly by 
  future space-borne gravitational wave experiments such as eLISA and NGO.
  In this paper, we show that the gravitational wave (GW) detectability strongly depends on
  the radial profile of the DM distribution.  So if the GW is
  detected, the power index, that is, the DM density distribution
  would be determined very accurately.  The DM density distribution
  obtained would make it clear how the IMBH has evolved from a seed
  black hole and whether the IMBH has experienced major mergers in the past.
  Unlike the $\gamma$-ray observations of DM annihilation,
  GW is just sensitive to the radial profile of the DM distribution
  and even to noninteracting DM.  Hence, the effect we demonstrate
  here can be used as a new and powerful probe into  DM properties.
\end{abstract}

\maketitle

\section{I. Introduction}
A large number of astrophysical and cosmological observations provide
convincing evidence for the existence of dark matter (DM).  The origin
and nature of DM remain largely unknown, and are among the most
challenging problems in current cosmology and most likely in particle
physics.

Recently, the distribution of DM around a black hole (BH) has
been under discussion in the context of indirect searches for DM
annihilation signals with $\gamma$-ray observations.  Gondolo and Silk
\cite{Gondolo:1999ef} first suggested that the adiabatic growth of a
BH creates a high density DM region, called the ``spike'', which
enhances the DM annihilation rate.  Subsequent work showed that the
existence of a DM spike around a supermassive black hole turns
out to be unlikely when one considers the effects of major merger
events of the host galaxies \cite{Merritt:2002vj}, off-center
formation of the seed BH \cite{Ullio:2001fb}, and scattering of dark
matter particles by surrounding stars
\cite{Merritt:2003qk,Bertone:2005hw}.  On the other hand, a DM
``minispike'' around an intermediate-mass black hole (IMBH), with a
mass range between $10^2$ and $10^6{\rm M}_\odot$, may survive if the
IMBH never experienced any major mergers
\cite{Zhao:2005zr,Bertone:2005xz}, as is expected to be the case for the many  IMBHs that have failed to merge into a supermassive BH.

The existence of such a spike structure is strongly dependent on the
details of BH formation and the history of major mergers, which are
far from clear.  In this paper, we propose that future gravitational
wave (GW) experiments can be used to probe the DM distribution around
BHs.  The existence of the dense DM region changes the gravitational
potential and affects the orbit of an object around the BH.  We
consider GWs from the coalescence event of a compact binary consisting
of a small mass object and an IMBH and evaluate the modification of
the GW signal by the existence of a DM minispike associated with the IMBH.
Such an event may be observed by  future space-based
interferometers such as the evolved Laser Interferometer Space Antenna (eLISA), the New Gravitational Wave Observatory (NGO) \cite{AmaroSeoane:2012km}, 
and DECi-hertz Interferometer Gravitational wave Observatory (DECIGO)
\cite{Kawamura:2011zz}.  We further discuss whether the eLISA-NGO
experiment is sensitive to the modification of the signal by the DM minispike.

Note that, while $\gamma$-ray observations can find the signal of DM
annihilation if DM is a weakly interacting massive particle, the
observation of GWs is just sensitive to the gravitational potential of
the DM halo and applicable even for noninteracting DM.
%for any type of DM.  
Therefore, future GW experiments offer a unique opportunity for
testing the existence of the DM spike around BHs.  Recently the GW
signatures of the DM has also been considered by Macedo $et \
al$.\cite{Macedo:2013qea}.

Let us describe the radial profile of the DM spike by a single
power law $\rho\propto r^{-\alpha}$ assuming a spherically symmetric
distribution of DM.  The adiabatic growth of the BH produces a dense
spike in the inner region of the minihalo within a radius of $r_{\rm
  sp}\sim 0.2 r_h$, where $r_h$ is the radius of gravitational
influence of the BH defined by $M(<r_h)=4\pi\int^{r_h}_0 \rho (r) r^2
dr=2M_{\rm BH}$, with $M_{\rm BH}$ being the BH mass
\cite{Merritt:2003qk}.  The final density profile of the spike depends
on the power law index $\alpha_{\rm ini}$ of the inner region of the
initial minihalo as $\alpha=(9-2\alpha_{\rm ini})/(4-\alpha_{\rm
  ini})$ \cite{Gondolo:1999ef,Quinlan:1994ed}.  If we assume the
Navarro, Frenk, and White profile \cite{Navarro:1995iw} for the
initial condition ($\alpha_{\rm ini}=1$), we get $\alpha=7/3$.  A Very
steep slope is generically predicted as we find $2.25<\alpha<2.5$ for
$0<\alpha_{\rm ini}<2$. Indeed,  
the largest plausible value of $\alpha$ 
corresponds to an initially isothermal darkmatter profile
$\alpha_{\rm ini}=2$. 
Le Delliou $et \ al$. \cite{LeDelliou} have given an analytic estimate of the radial
distribution of the profile. To be conservative, 
we restrict $\alpha$ to below 3 throughout this paper.

In summary, in this paper, we assume the DM distribution of a
minispike is described by
\begin{align}
 \rho(
%{\boldsymbol{x}}
r
)&= \rho_{\rm sp}\left(\frac{r_{\rm
 sp}}{r}\right)^{\alpha}
\,\,\,(r_{\rm min} \le r \le r_{\rm sp}),
\label{Eq:rhoDM}
\end{align}
where $\rho_{\rm sp}$ is the normalization of the DM density.  For an
IMBH with the mass of $M_{\rm BH}=10^3 M_{\odot}$ and the total mass
of the DM minihalo of $M_{\rm halo}=10^6 M_{\odot}$, we get
$\rho_{\rm sp} = 379 \ M_{\odot}/\text{pc}^{3}$ and $r_{\rm sp} =
0.33 \text{pc}$.  Beyond the spike radius $r_{\rm sp}$, the DM
distribution obeys the NFW profile with the concentration parameter
$c=6.6$ estimated based on the fitting formula given, e.g., in
\cite{Duffy:2008pz}.  The minimum radial distance is taken to be
$r_{\text{min}}= r_{\text{ISCO}}$ where $r_{\text{ISCO}}$ is the
innermost stable circular orbit (ISCO) given by $r_{\text{min}} =
r_{\text{ISCO}} = 6GM_{\text{BH}}/c^2$.

\section{II. Formulation}
\subsection{GWs from binary inspiral}
Let us consider gravitational waves from a binary system consisting of
an IMBH with a mass of $M_{\text{BH}} \sim 10^3 M_{\odot}$ and a
compact object with a mass of $\mu \sim 1 M_{\odot}$.  For simplicity,
we make the following idealization.  First, we treat the star as a
test particle and we call it a ``particle'' in the following.  Second,
we assume that the DM density is unperturbed even when the star orbits
in the DM \textcolor{red}{minispike}. Gravitational heating of the DM minispike due to the
particle may be noticeable within the Hill sphere of the
particle because of the gravity of the central IMBH.  In the case of
our $1\msol$-$10^3\msol$ binary, the Hill radius is 10\% of the
orbital radius and we ignore possible heating effects in the first
order approximation.  Then, the equation of motion for the particle is
written as
\begin{align}
\frac{d^2 r}{dt^2} &= - \frac{GM_{\text{eff}}}{r^2} - \frac{F}{r^{\alpha -1}} + \frac{l^2}{r^3}, 
\label{Eq:EOM}
\end{align}
where $l$ is the angular momentum of the particle per its mass, and
$M_{\text{eff}}$ and $F$ are
\begin{align}
&
\begin{array}{ccr}
\meff= \left\{
\begin{array}{l} 
%\vspace{0.5em}
%\mbh +
% \dfrac{4\pi\rsp^{\alpha}\rhosp}{3-\alpha}\left(\rsp^{3-\alpha}-\rmin^{3-\alpha}\right) \\
\vspace{0.5em}
\mbh - \dfrac{4\pi\rsp^{\alpha}\rhosp}{3-\alpha} \rmin^{3-\alpha}, \\
\mbh,
\end{array} 
\right.
&
F=\left\{
\begin{array}{c}
% \vspace{0.5em}
% 0  \\
\vspace{0.5em}
\dfrac{4\pi G\rsp^{\alpha}\rhosp}{3-\alpha} \\
0
\end{array}
 \right.
&
\begin{array}{c}
%\vspace{0.5em}
%(r\le r \le \rsp) \\
\vspace{0.5em}
(\rmin \le r \le \rsp), \\
(r < \rmin).
\end{array} 
\end{array}
\end{align}

In the first term of the right-hand side of Eq. (\ref{Eq:EOM}), the DM
\textcolor{red}{minispike} modifies the effective mass of the central IMBH.
%\textcolor{red}{
%At $r_{\text{ISCO}}$ there is a transition from inspiral to a final plunge.
% The DM mini-halo dose not exist stably within this boundary.
% So the BH mass seems as reduced by the DM mass within $r_{\text{ISCO}}$.
%}
  The second term contains information of the DM \textcolor{red}{minispike} radial distribution.  The
third term represents a centrifugal force.  
Note that the DM particles do not exist stably within $r_{\rm
    min}=r_{\text{ISCO}}$ and we assume $\rho=0$.  For $\rmin \le r
  \le \rsp$, $F$ represents the effect of DM assuming that the DM
  distribution is given by Eq. (\ref{Eq:rhoDM}) for $0 \le r \le
  \rsp$.  Instead, the effective mass of the BH $\meff$ is reduced to
  offset the extra mass in $0 < r < \rmin$.

If we assume that the second term is much smaller than the first term,
\begin{align*}
\varepsilon \left(\dfrac{r}{r_{\text{min}}}\right)^{3-\alpha} \ll 1 \ \ \ \ \left(\varepsilon \equiv \dfrac{Fr_{\text{min}}^{3-\alpha}}{GM_{\text{eff}}} \right), 
\end{align*}
we can treat the term which involves information on the DM \textcolor{red}{minispike} as a
perturbation and expand equations in powers of $\varepsilon$, which
is a dimensionless parameter depending on the power index $\alpha$.

When the particle stably orbits around the IMBH at a constant radius
$R$, the left-hand side of the equation of motion vanishes.  In this
case, the GW waveforms are given by
\begin{align}
 h_{+} &= \dfrac{1}{r} \dfrac{4G\mu \omega_s^2 R^2}{c^4} \dfrac{1 + \cos
 ^2 \iota}{2} \cos\left( 2\omega_s t\right),
 \label{Eq:waveform1}\\
 h_{\times} &= \dfrac{1}{r} \dfrac{4G\mu \omega_s^2 R^2}{c^4}  \cos \iota \sin\left( 2\omega_s t\right)
\label{Eq:waveform2}
\end{align}
to the lowest order approximation where $\iota$ is the inclination
which is the angle between the normal to the orbit and the
line of sight, and $2\omega_s$ is the GW frequency.

\subsection{Waveforms including GW back-reaction}
Next, we include the effect of the GW backreaction within the
linearized theory of  Einstein's general relativity.  The
orbital radius and frequency are no longer constant, because GW
radiation energy $E_{\text{GW}}$ is taken from the orbital energy
$E_{\text{orbit}}$ of the particle.  The relation between the orbital
radius $R$ and the time $t$ is given by the energy balance (e.g., Chap. 4
of \cite{Maggiore}),
\begin{align}
& \dfrac{dE_{\text{orbit}}}{dt} = - \dfrac{d E_{\text{GW}}}{dt},
\end{align}
where 
\begin{align}
 \dfrac{dE_{\text{orbit}}}{dt} &= \left( \dfrac{G\mu M_{\text{eff}}}{2R^2} + \dfrac{4-\alpha}{2}\dfrac{\mu F}{R^{\alpha-1}} \right) \dfrac{dR}{dt}\\
 \dfrac{dE_{\text{GW}}}{dt} &= \dfrac{32}{5}\dfrac{G\mu^2}{c^5}R^4 \omega_s^6.
\end{align}
Using this relation, we can compute the orbital frequency $\omega_s$
and $R$ as a function of time.  To include the GW backreaction in the
GW waveforms, we replace the constant parameters $\omega_s$ and $R$ in
Eqs. (\ref{Eq:waveform1}) and (\ref{Eq:waveform2}) by time-dependent
functions $\omega_s(t)$ and $R(t)$.  Then, we perform the Fourier transform 
$\tilde{h}(f) = \int h(t) \exp(i2\pi ft) dt$ to compare the theoretical waveforms with GW experiments.  
The stationary phase approximation enables us to obtain the GW
waveforms in Fourier space expanded in $\varepsilon$ (e.g., Chap. 4 of \cite{Maggiore}),

%The stationary phase approximation then enables us to obtain the GW
%waveforms in Fourier space expanded in $\varepsilon$,

%\textcolor{red}{  Given a time function $B(t)=A(t)\cos\phi(t)$
%where $d\ln A/dt\ll d\phi/dt$ and $d^2\phi/dt^2\ll(d\phi/dt)^2$, the stationary phase approximation
%enable us to obtain the Fourier transform $\tilde{B}(f)$ in $f\geq 0$,
%\begin{align}
% \tilde{B}(f)=\dfrac{1}{2}A(t)\left(\dfrac{df}{dt}\right)^{-\frac{1}{2}}e^{i\left(2\pi ft-\phi(t)-\frac{\pi}{4}\right)},
%\end{align}
%where $t$ is defined as $2\pi f=d\phi/dt$ . Using this approximation, we get the GW waveforms in Fourier space expanded in
%$\varepsilon$,}

\begin{align}
  \tilde{h}_{+} \left( f \right) &= \left(\dfrac{5}{24} \right)^{1/2} \frac{e^{i\Psi \left(f\right)  }}{\pi^{2/3}f^{7/6}}
\dfrac{c}{r} \left( \dfrac{GM_c}{c^3}\right)^{5/6}  \dfrac{1+\cos^2\iota}{2}  \left[ 1+   \dfrac{7-2\alpha}{3}  
\left( \dfrac{GM_{\text{eff}}}{\pi^2 r_{\text{min}}^3 f^2}\right)^{\left(3-\alpha\right)/3}\varepsilon  + \cdots \right] ,  \\
  \tilde{h}_{\times} \left( f \right) &= \left(\dfrac{5}{24}
 \right)^{1/2} \dfrac{i e^{i\Psi \left(f\right)  }}{\pi^{2/3}f^{7/6}}  \dfrac{c}{r}
 \left( \dfrac{GM_c}{c^3}\right)^{5/6} \cos\iota \left[ 1+   \dfrac{7-2\alpha}{3} 
 \left( \dfrac{GM_{\text{eff}}}{\pi^2 r_{\text{min}}^3 f^2}\right)^{\left(3-\alpha\right)/3}\varepsilon  + \cdots \right], \\
  \Psi &= 2\pi f\left(t_c + \dfrac{r}{c}\right) - \Phi_0 -\dfrac{\pi}{4}
 + 2\left( \dfrac{GM_c}{c^3} 8\pi f \right)^{-5/3} + \Delta\Psi,
 \label{Eq:phase} 
\intertext{with}
\Delta \Psi &= 2\left( \dfrac{GM_c}{c^3} 8\pi f \right)^{-5/3}
 \left[ \dfrac{10}{3}\dfrac{2\alpha-5}{2\alpha-11} \left(
 \dfrac{GM_{\text{eff}}}{\pi^2 r_{\text{min}}^3
 f^2}\right)^{\left(3-\alpha\right)/3} \varepsilon
 -\dfrac{5}{9}\dfrac{\left(2\alpha-1\right)\left(4\alpha-11\right)}{4\alpha-17}
 \left( \dfrac{GM_{\text{eff}}}{\pi^2 r_{\text{min}}^3
 f^2}\right)^{\left[2\left(3-\alpha\right)\right]/3} \varepsilon^2 + \cdots  \right],  
\label{Eq:DeltaPsi}
\end{align}
where $t_c$ is the value of retarded time at coalescence, $\Phi_0$ is
the value of the phase at coalescence, $M_c = \mu^{3/5}
M_{\text{eff}}^{2/5}$ is the chirp mass, and $\Psi = \int 2 \omega_s (t) dt$
 is the phase of the GW waveform.  These expansions are valid for the frequency $f$ for
which higher order terms are negligible.

\begin{figure}
\begin{center}
\includegraphics[width=10cm,clip,bb=50 50 230 176]{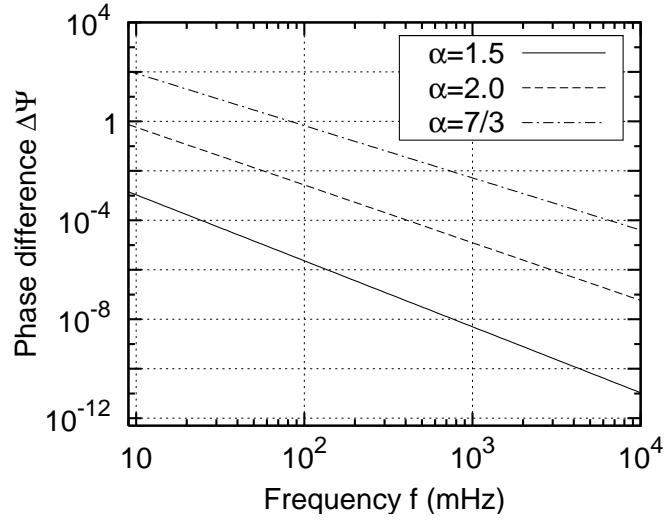}
\caption{Phase difference $\Delta \Psi$ against frequency. Solid line
  is for $\alpha = 1.5$, dashed line is for $\alpha = 2.0$, and
  dot-dashed line is for $\alpha = 7/3$.}
\label{Fig:Phase}
\end{center}
\end{figure}

\begin{figure}
\begin{center}
\includegraphics[width=10cm,clip,bb=50 50 230 176]{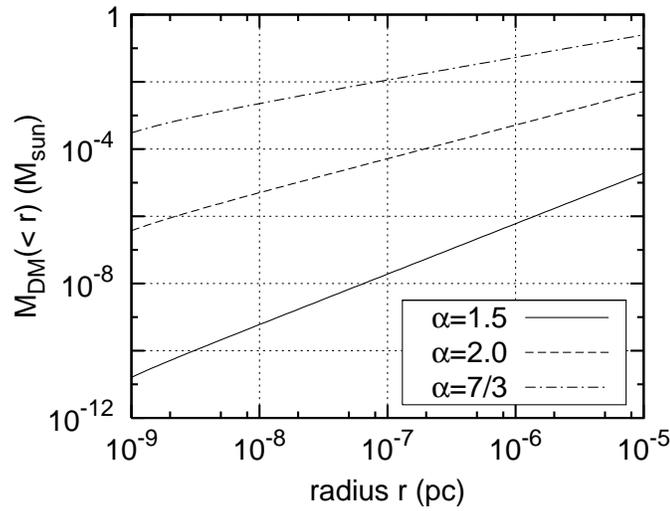}
\end{center}
\caption{Mass of DM \textcolor{red}{minispike} within orbital radius $r$. A steeper density
  distribution contains more DM mass within the radius $r$. The solid line
  is $\alpha = 1.5$, the dashed line is $\alpha = 2.0$, and the dot-dashed line
  is $\alpha = 7/3$.}
\label{Fig:Mass}
\end{figure}

In Eq. (\ref{Eq:phase}), the phase of the GW is modified by the
presence of the DM, which is expanded in powers of $\varepsilon$.
Since GW interferometers are very sensitive to the phase of the
signal, this phase difference is crucial for distinguishing the
existence of the DM minispike.  In Fig. \ref{Fig:Phase}, we plot the phase
difference $\Delta\Psi$ caused by the DM \textcolor{red}{minispike}, taking into account terms up
to  second order in $\varepsilon$.  We see $\Delta \Psi$ increases
for low frequencies and for large $\alpha$.  This can be explained by
the fact that the orbit of the object is affected only by the DM mass
inside the orbital radius.  More phase difference is produced when the
inner mass is large.  As shown in Fig. \ref{Fig:Mass}, the enclosed DM
mass increases as the radius or $\alpha$ increases.  Since a low
frequency of the GW corresponds to a large orbital radius, a large phase
difference is produced at low frequencies.  A larger value of $\alpha$
, or equivalently a steeper density distribution, leads to a larger inner mass,
which also results in a larger phase difference.

\section{III. Observation of GWs}
\subsection{Matched filtering}
Let us discuss whether or not this effect is testable by future GW experiments.
The search for GW signals is performed by matched filtering analysis,
in which one correlates detector output with theoretical template.
The signal-to-noise ratio obtained in the matched filtering technique is defined by
\begin{align}
&\left( \dfrac{S}{N} \right)^2 = \dfrac{\displaystyle \left[ \int_{f_{\text{ini}}}^{\infty}df \ \dfrac{\tilde{h}\left(f\right)\tilde{h}_t^{\ast} \left(f\right)+ \tilde{h}^{\ast}\left(f\right)\tilde{h}_t\left(f\right) }{  S \left(f\right)} \right]^2 }{ \displaystyle\int_{f_{\text{ini}}}^{\infty} df \ \dfrac{\left| \tilde{h}_t \left(f\right)\right|^2 }{ S \left(f\right)}},  
\label{Eq:SN}
\end{align}
where $\tilde{h}(f)$ is the GW signal coming to the detector,
$\tilde{h}_t(f)$ is the template, $S(f)$ is the spectral density of
the detector noise, and $f_{\text{ini}}$ is the frequency of the
inspiral GW when the observation started.  In the following example,
we assume the eLISA experiment, whose noise spectrum is given in
Ref. \cite{AmaroSeoane:2012km}.

In Eq. \eqref{Eq:SN}, the numerator is a noise-weighted correlation
between the template and the true signal, and the denominator is the
renormalization factor.  When the template matches the true waveform,
$S/N$ is maximized.  Thus, $S/N$ is an indicator to tell us whether the
waveform of the template is present in the detector or not.
\textcolor{red}{
We would claim detection of GW if the associated $S/N$ ratio is larger than the 
predefined threshold value.
%This threshold depends on, for instance, detection method, noise properties and the number of detectors used.
In the literature, e.g., Ref. \cite{PRD78_042002}, $S/N > 8$ is required to claim detection.
}

\subsection{Detectability of the effect of a DM \textcolor{red}{minispike} around an IMBH}
Let us consider an observation of GWs from the $1 \msol$ particle
inspiraling into the $10^3\msol$ IMBH, which would be detectable by
the eLISA experiment.  We assume this binary is surrounded by a DM
mini-spike whose distribution is given by Eq. (\ref{Eq:rhoDM}).  In
this setup, a frequency integration from $f_{\text{ini}}=22.7 \ 
\text{mHz}$ corresponds to a $5.0$ yr observation until the
coalescence (corresponding roughly to the expected eLISA frequency
band and observation time). Note that when $f_{\rm ini}=22.7$mHz, the
particle is about $10^{-8}$ pc away from the IMBH and well within the
minispike.

\begin{figure}
\begin{center}
\includegraphics[width=10cm,clip,bb=50 50 230 176]{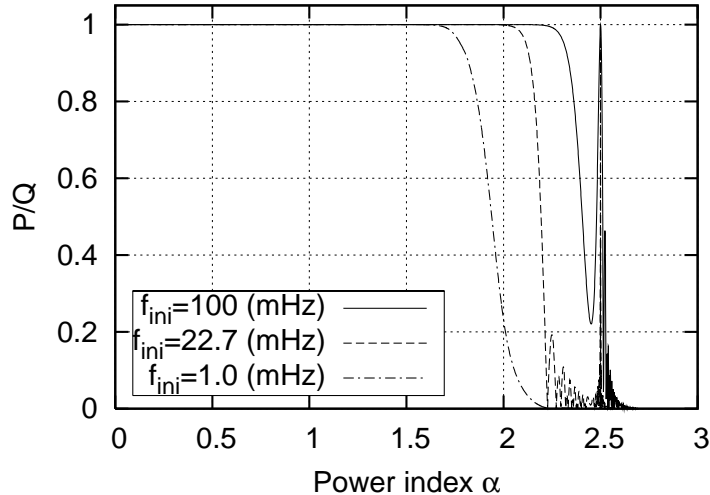}
\caption{$P/Q$ against power index $\alpha$. Three different curves
  show $P/Q$ for three different values of initial frequency
  $f_{\text{ini}}$, namely %differential 
different
observation time. The solid
  line is for $f_{\text{ini}}=100 \ \text{mHz}$, the dashed line is for
  $f_{\text{ini}}=22.7 \ \text{mHz}$ and the dot-dashed line is for
  $f_{\text{ini}}=1.0 \ \text{mHz}$. }
\label{Fig:PQ}
\end{center}
\end{figure}

In Fig. \ref{Fig:PQ}, we show how much the $S/N$ is degraded when one
applies a template predicted without considering the DM effect on the
signal with the DM effect.  The vertical axis represents a degradation
rate $P/Q$, where $P$ is $S/N$ calculated assuming a template of a
waveform without the DM effect ($\varepsilon \rightarrow 0$) and $Q$
is $S/N$ calculated with a template including the DM effect up to the
second order.  If the effect of the DM is small, there is little
difference between the two templates and $P/Q$ becomes $1$.
Conversely, if DM potential induces significant phase difference, the
value of $P$ decreases, since the template and the signal have less
correlation.

As discussed in the previous section, the phase difference becomes
significant for large $\alpha$, and, from Fig. \ref{Fig:PQ}, we find
$P/Q$ largely deviates from $1$ for $\alpha\gtrsim 2$.  
%This result
%indicates that GW observation can distinguish whether a DM halo of
%$\alpha\gtrsim 2$ exists around the IMBH.  
\textcolor{red}{This means that} in order to extract
inspiral signals under the effect of a DM \textcolor{red}{minispike}, we must prepare
templates including the DM effect \textcolor{red}{when $\alpha\gtrsim 2$}.  
\textcolor{red}{For example, a GW signal that gives $S/N = 8$ when
we use the correct template would then give $S/N = 0.8$ if we use the incorrect
one and $P/Q = 0.1$. 
We miss this signal if we do not take account of the effect of a DM minispike.
%In any case, our discussion with P/Q is applicable for any value of S/N
%ratio.
This result \textcolor{red}{in turn}  
indicates that GW observation can distinguish whether a DM minispike of
$\alpha\gtrsim 2$ exists around the IMBH.  
}

In Fig. \ref{Fig:PQ}, we also plot
the cases for different initial integration frequencies, which
corresponds to different observation time.  Since the phase difference
becomes larger at low frequency, $P/Q$ is suppressed for smaller
value of $\alpha$ when one observes a longer time period.  The peak
seen at $\alpha=2.5$ originates from the zero crossing of the first
term of Eq. (\ref{Eq:DeltaPsi}).

\section{IV. Conclusion}
\textcolor{red}{It has been expected that $\gamma$-ray and/or neutrino
  observations on DM halos enable us to study properties of
  annihilating DM particles (e.g., Ref. \cite{APJ761_91A_2012}).
  %This is true only if DM particles are annihilating.
  In this paper, we proposed a new method to explore a dense DM
  minihalo, the so-called DM minispike, using GW signals that is more
  powerful when DM does not annihilate.  Namely, we have demonstrated
  a method to probe the DM distribution around an IMBH by using GW
  direct detection experiments.  Considering a GW signal from a
  compact object inspiraling into an IMBH, we have computed how the
  GW waveform is modified by the gravitational potential of the DM
  halo.  Thanks to the fact that a GW interferometer measures the
  phase of the signal with very good accuracy, we found that a GW
  experiment such as eLISA and NGO is sensitive to the phase shift caused
  by the DM potential.  Indeed, we found that the GW observation can
  detect the DM effect when DM does not annihilate and its profile is
  steep enough.
  % when DM does not annihilate and its profile is steep enough, GW
  % observation may provide a new observational probe for the DM
  % distribution near BHs and would be helpful for testing the
  % existence of a dense DM minihalo, the so-called DM spike.
  Therefore, GW observation would be a complementary method for
  testing the existence of a DM distribution: while $\gamma$-ray and/or neutrino
  observations are powerful to probe annihilating DM particles, the GW
  test offers a unique opportunity to detect the presence of
  nonannihilating DM.  This may even offer hints to the formation
  history of BHs, since formation of DM spikes strongly depends on how
  BHs evolved.}

%We have demonstrated a method to probe the DM distribution around an
%IMBH by using GW direct detection experiments.  Considering a GW
%inspiral signal from inspiral of a compact object falling in an IMBH,
%we have computed how the GW waveform is modified by the gravitational
%potential of the DM halo.  Thanks to the fact that a GW interferometer
%measures the phase of the signal with very good accuracy, we found
%that a GW experiment such as eLISA/NGO is sensitive to the phase
%shift caused by the DM potential.  

%Accordingly, \textcolor{red}{ when DM does not annihilate and its density
%becomes steep enough,} GW observation may
%provide a new observational probe for the DM distribution near BHs and
%would be helpful for testing the existence of a dense DM minihalo, the
%so-called DM spike.  
%This may even offer hints to the formation history of
%BHs, since formation of DM spikes strongly depends on how BHs evolved.
%\textcolor{red}{
%On the other hand, 
%if DM is annihilating, forming a core and the DM density profile becomes
%flat, we can not infer an existence of a DM halo from GW signal. 
%In this case, a DM halo would be better searched using gamma-ray or
%neutrino observations. 
%}

In future work, we plan to extend the investigation for different
values of mass distribution parameters, such as the power index
$\alpha$ and the mass of the compact objects and the halo.  We will also
estimate to what degree future GW experiments can determine the mass
distribution by computing expected errors on the parameters.

%\textcolor{magenta}{Finally we remark on a recent work which studies 
%the GW signatures of the DM \cite{Macedo:2013qea}. They have
%considered a massive but compact self-gravitating configuration of DM
%particles. A smaller stellar mass object then spiralls into the center
%of the configration. The differences between their work and  the
%current one arise due to the fact that they do not consider any central
%compact object inside the DM configuration while the central IMBH with
%in the DM halo is the important ingradient of our setting.   
%}

%Finally we remark on the dynamical friction effect.  In this paper, we
%have ignored dynamical friction.  Even if GW back reaction is
%ignored, the dynamical friction alone makes the smaller compact object
%infall into the IMBH.  Using the Chandrasekhar formula (e.g.,
%\cite{GalacticDynamics}), we found however that this effect produces
%negligible phase difference when computing the SN degradation rate.

\begin{acknowledgments}
  We thank Jun'ichi Yokoyama for his valuable comments.  This work is
  supported by JSPS KAKENHI Grant No. 23340058 and No. 25103504.
  %supported by the Grant-in-Aid for Scientific Research (B) No. 23340058
  %from the Japan Society for the Promotion of Science.  
  S.K. is partially supported by the Grant-in-Aid for Scientific Research 
  No. 24740149.  The work of J.S. is supported in part by ERC Project
  No. 267117 (DARK) hosted by Universit\'e Pierre et Marie Curie - Paris 6.
\end{acknowledgments}


\begin{thebibliography}{100}

 
%\cite{Gondolo:1999ef}
\bibitem{Gondolo:1999ef} 
  P.~Gondolo and J.~Silk,
  %``Dark matter annihilation at the galactic center,''
  Phys.\ Rev.\ Lett.\  {\bf 83}, 1719 (1999).
  %[astro-ph/9906391].
  %%CITATION = ASTRO-PH/9906391;%%


%\cite{Merritt:2002vj}
\bibitem{Merritt:2002vj} 
  D.~Merritt, M.~Milosavljevic, L.~Verde and R.~Jimenez,
  %``Dark matter spikes and annihilation radiation from the galactic center,''
  Phys.\ Rev.\ Lett.\  {\bf 88}, 191301 (2002).
  %[astro-ph/0201376].
  %%CITATION = ASTRO-PH/0201376;%%


%\cite{Ullio:2001fb}
\bibitem{Ullio:2001fb} 
  P.~Ullio, H.S.~Zhao and M.~Kamionkowski,
  %``A Dark matter spike at the galactic center?,''
  Phys.\ Rev.\ D {\bf 64}, 043504 (2001).
  %[astro-ph/0101481].
  %%CITATION = ASTRO-PH/0101481;%%



%\cite{Merritt:2003qk}
\bibitem{Merritt:2003qk} 
  D.~Merritt,
  %``Evolution of the dark matter distribution at the galactic center,''
  Phys.\ Rev.\ Lett.\  {\bf 92}, 201304 (2004).
  %[astro-ph/0311594].
  %%CITATION = ASTRO-PH/0311594;%%

%\cite{Bertone:2005hw}
\bibitem{Bertone:2005hw} 
  G.~Bertone and D.~Merritt,
  %``Time-dependent models for dark matter at the Galactic Center,''
  Phys.\ Rev.\ D {\bf 72}, 103502 (2005).
  %[astro-ph/0501555].
  %%CITATION = ASTRO-PH/0501555;%%

%\cite{Zhao:2005zr}
\bibitem{Zhao:2005zr} 
  H.~-S.~Zhao and J.~Silk,
  %``Mini-dark halos with intermediate mass black holes,''
  Phys.\ Rev.\ Lett.\  {\bf 95}, 011301 (2005).
  %[astro-ph/0501625].
  %%CITATION = ASTRO-PH/0501625;%%

%\cite{Bertone:2005xz}
\bibitem{Bertone:2005xz} 
  G.~Bertone, A.~R.~Zentner, and J.~Silk,
  %``A new signature of dark matter annihilations: gamma-rays from intermediate-mass black holes,''
  Phys.\ Rev.\ D {\bf 72}, 103517 (2005).
  %[astro-ph/0509565].
  %%CITATION = ASTRO-PH/0509565;%%


%\cite{AmaroSeoane:2012km}
\bibitem{AmaroSeoane:2012km} 
  P.~Amaro-Seoane, S.~Aoudia, S.~Babak, P.~Binetruy, E.~Berti, A.~Bohe, C.~Caprini, M.~Colpi, {\it et al.},
  %``eLISA: Astrophysics and cosmology in the millihertz regime,''
  arXiv:1201.3621. % [astro-ph.CO].
  %%CITATION = ARXIV:1201.3621;%%

%\cite{Kawamura:2011zz}
\bibitem{Kawamura:2011zz} 
  S.~Kawamura, M.~Ando, N.~Seto, S.~Sato, T.~Nakamura, K.~Tsubono, N.~Kanda, T.~Tanaka {\it et al.},
  %``The Japanese space gravitational wave antenna: DECIGO,''
  Classical Quantum Gravity  {\bf 28}, 094011 (2011).
  %%CITATION = CQGRD,28,094011;%%

%\cite{Macedo:2013qea}
\bibitem{Macedo:2013qea} 
  C.~F.~B.~Macedo, P.~Pani, V.~Cardoso, and L.~C.~B.~Crispino,
  %``Into the lair: gravitational-wave signatures of dark matter,''  arXiv:1302.2646 [gr-qc].  %%CITATION = ARXIV:1302.2646;%%  %1 citations counted in INSPIRE as of 12 Apr 2013
  arXiv:1302.2646.% [gr-qc].


%\cite{Navarro:1995iw}
\bibitem{Navarro:1995iw} 
  J.~F.~Navarro, C.~S.~Frenk, and S.~D.~M.~White,
  %``The Structure of cold dark matter halos,''
  Astrophys.\ J.\  {\bf 462}, 563 (1996).
  %[astro-ph/9508025].
  %%CITATION = ASTRO-PH/9508025;%%


\bibitem{LeDelliou} 
M.~Le~Delliou, R.~N.~Henriksen, and J.~D.~MacMillan, 
%Astronomy and Astrophysics, {\bf 526} A13 (2011)
Astron. Astrophys. {\bf 526}, A13 (2011).
%arXiv:0911.2238 [astro-ph.GA].




%\cite{Duffy:2008pz}
\bibitem{Duffy:2008pz} 
  A.~R.~Duffy, S.~Joop, S.~T.~Kay, and C.~Dalla~Vecchia,
  %``Dark matter halo concentrations in the Wilkinson Microwave Anisotropy Probe year 5 cosmology,''
  Mon.\  Not.\ R.\ Astron.\ Soc.\  {\bf 390}, L64 (2008).
  %[astro-ph/0804.2486].
  %%CITATION = ASTRO-PH/0804.2486;%%

%\cite{Quinlan:1994ed}
\bibitem{Quinlan:1994ed} 
  G.~D.~Quinlan, L.~Hernquist, and S.~Sigurdsson,
  %``Models of Galaxies with Central Black Holes: Adiabatic Growth in Spherical Galaxies,''
  Astrophys.\ J.\  {\bf 440}, 554 (1995).
  %[astro-ph/9407005].
  %%CITATION = ASTRO-PH/9407005;%%

\bibitem{Maggiore}
M.~Maggiore, $Gravitational \ Waves: \ Theory \ and \ Experiments$
	(Oxford University Press, 2007), Vol. 1.


\bibitem{PRD78_042002}
B.~Abbott {\it et. al.} (LIGO Scientific Collaboration)
%''Search of S3 LIGO data for gravitational wave signals from spinning
%black hole and neutron star binary inspirals''
Phys. Rev. D {\bf 78}, 042002 (2008).



\bibitem{APJ761_91A_2012}
M.~Ackermann {\it et. al.} (Fermi-LAT Collaboration),
%''Constraints on the Galactic Halo Dark Matter from Fermi-LAT Diffuse Measurements''
  Astrophys.\ J.\  {\bf 761}, 91A (2012).










%\bibitem{GalacticDynamics}
%J.~Binney and S.~Tremaine ``Galactic Dynamics'' (2nd ed.) Princeton University
%	Press (2008)

\end{thebibliography}
\end{document}